\documentclass[11pt]{article}

\usepackage{geometry} 
\usepackage{graphicx} 

\usepackage{booktabs} 
\usepackage{array} 
\usepackage{paralist} 
\usepackage{verbatim} 
\usepackage{subfig} 
\usepackage{amsmath} 
\usepackage{amssymb} 
\usepackage{hyperref} 
\usepackage{epstopdf} 
\usepackage{bbm}
\usepackage{color}

\usepackage{fancyhdr} 
\usepackage{sectsty}
\allsectionsfont{\sffamily\mdseries\upshape} 

\usepackage[nottoc,notlof,notlot]{tocbibind} 
\usepackage[titles,subfigure]{tocloft} 

\title{Quantitative analysis of the evolution of novelty in cinema through crowdsourced keywords}
\author{Sameet Sreenivasan$^{1,2,3}$}
\begin{document}

\maketitle

\begin{flushleft}
{\bf 1} Social and Cognitive Networks Academic Research Center, Rensselaer Polytechnic Institute, 110 8$^{th}$ Street, Troy, NY, 12180, USA
\\
{\bf 2} Department of Computer Science, Rensselaer Polytechnic Institute, 110 8$^{th}$ Street, Troy, NY, 12180, USA
\\
{\bf 3} Department of Physics, Rensselaer Polytechnic Institute, 110 8$^{th}$ Street, Troy, NY, 12180, USA
\\
E-mail: sreens@rpi.edu
\end{flushleft}

\section*{Abstract}
The generation of novelty is central to any creative endeavor. Novelty generation and the relationship between novelty and individual  hedonic value have long been subjects of study in social psychology. However, few studies have utilized large-scale datasets to quantitatively investigate these issues. Here we consider the domain of American cinema and explore these questions using a database of films spanning a $70$ year period. We use crowdsourced keywords from the Internet Movie Database as a window into the contents of films, and prescribe novelty scores for each film based on occurrence probabilities of individual keywords and keyword-pairs. These scores provide revealing insights into the dynamics of novelty in cinema. We investigate how novelty influences the revenue generated by a film, and find a relationship that resembles the Wundt-Berlyne curve. We also study the statistics of keyword occurrence and the aggregate distribution of keywords over a $100$ year period.

\section*{Introduction}

Over the last century, cinema has carved out an indelible niche in human culture, and filmmaking has come to be regarded as an art-form its own right. The film industry of the United States in particular, has had a major influence on the evolution of cinema over the course of its history, and is currently the third largest producer of films in the world, with a global audience and a gross turnover averaging $29.5$ billion US dollars over the last five years reported \cite{mpaa}. Despite the fact that trends associated with films, the dissection of their respective successes and  failures, and their individual artistic merit are all subjects of avid debate and discussion in the public realm, and although the economics of film has been extensively researched \cite{Devanybook}, no studies, to our knowledge, have quantitatively analyzed the large scale features of novelty in film plots and the patterns associated with their evolution. With the advent of culturomics as an emerging science \cite{Michel}, it is natural to attempt to bridge this gap with the aid of comprehensive sources of film data such as the Internet Movie Database (IMDb).

The Internet Movie Database (www.imdb.com) is a comprehensive online database containing information on films, television programs and videogames which, according to the site, has ``more than $100$ million data items including more than $2$ million movies''. This in large part is made possible by allowing registered users of the site to add new database items or edit the information associated with existing ones. One such category of user-generated information at the center of this study, is that of {\it plot-keywords} consisting of single words, or word-strings associated with each item. If a keyword proposed by a user is semantically close to a keyword that already exists (i.e., has already been created for association with one or more films), then the user is prompted to use the existing keyword, thus suppressing the creation of synonymous keywords. In the context of films, keywords describe any of a number of aspects of film including but not limited to thematic plot-elements ({\it father-son-relationship}, {\it power}, {\it fame}), specific story elements ({\it tied-to-a-chair}, {\it held-at-gunpoint}, {\it breaking-and-entering}), location references ({\it manhattan-new-york-city}, {\it coffee-shop}, {\it Chevron-gas-station}) specific visual or object references ({\it life-magazine}, {\it characters-point-of-view-camera-shot}, {\it coin-flipping-in-the-air}) or high-level features of the film ({\it independent-film}, {\it female-nudity}, {\it cult-film}). Plot-keywords are thus qualitative descriptors spanning several scales of detail and specificity, and they potentially constitute a rich information set  capable of yielding valuable insights into the evolution of films over time.

The dynamics of tagging - the process of users contributing keywords to associate with specific items - as well as folksonomy - the classification of items based on these collective tags - have been widely studied in the context of blogs, photo-sharing and social bookmarking \cite{Halpin,Cattuto,Brooks,McAuley,Farooq,Levy,Shepitsen,Szomszor,Gupta,Milicevic}. A general consensus derived from these studies is that despite a lack of central control, shared vocabularies with stable probability distributions over words emerge as a result of collaborative tagging. For example, Halpin et al. \cite{Halpin} showed that the relationship between the frequency of a tag's usage and its rank (based on how frequently it is used) is a power-law, and further proposed a model for tagging dynamics based on preferential attachment that could yield such a relationship. Almost concurrently, Cattuto et al. \cite{Cattuto} showed that the frequency-rank plot for tags obtained from {\it Del.icio.us} and {\it Connotea} indicated a power-law relationship, and demonstrated that a Yule-Simon model with long-term memory for tagging dynamics could yield this relationship. In the context of information retrieval, Levy and Sandler \cite{Levy} showed how social tags associated with musical tracks (on a {\it Last.fm} dataset) defined a semantic space that could enable efficient mood-based clustering and retrieval. Similarly, there have been several studies \cite{Shepitsen,Szomszor, Gupta, Milicevic} that have focussed on exploring the use of tags for personalized recommendation and query based retrieval. As a representative example, Szomnsor et al, \cite{Szomszor} investigated the extent to which combining tags obtained from IMDb and ratings data obtained from Netflix could generate better taste profiles for users, and thus yield a predictor of their ratings for an unseen film. Finally, although unrelated to social tagging but still within the larger domain of collaborative editing, Mesty\'{a}n et al. \cite{Mestyan} showed how user activity data on Wikipedia pertaining to a particular film's entry could yield an early predictor for the box-office success of the film.

In contrast to the above studies, the motivation of this work is to utilize the IMDb plot-keywords dataset as a window into the evolution of films and their content over the course of the last century, and in the process investigate certain aspects of novelty generation in the arts. The characterization of novelty, and the processes that lead to it, have been subjects of thorough investigation in psychology and social science \cite{Koestler, Boden,Hofstadter}. Several of these studies emphasize the role of the combinational process - one that combines existing ideas in a manner not encountered earlier - in novelty generation, in contrast to the process of introducing fundamentally new concepts from scratch.
Another aspect of sustained research interest \cite{Sluckin,Berns,Anand,Saunders} is the relationship between the novelty of an item and the hedonic value (or pleasure) derived by an individual upon its consumption. The standard paradigm here, resulting from the pioneering work of Wundt \cite{Wundt} and Berlyne \cite{Berlyne}, is captured by the Wundt-Berlyne curve, which posits that increasing novelty initially results in increasing hedonic value until it reaches a maximum. Further increasing novelty beyond this intermediate level results in a rapid decline in hedonic value. In summary, the red inverted-U shaped Wundt-Berlyne curve posits that individuals seek a balance between familiarity and novelty, shying away from the banal as strongly as from the radically unfamiliar. 


The issues of novelty creation and novelty optimization are undoubtedly relevant to the business of cinema. A significant portion of film criticism, commentary and discussion is devoted to analyzing the novelty in the writing and execution of film plots. In addition, one among the various factors responsible in successfully securing the financing and distribution of a film, is its conformity to current trends and past conventions. However, little is known in a quantitative sense regarding the degree to which the competing objectives of novelty and conformity are balanced in the process of new content creation. The plot-keywords dataset has the potential to serve as a starting point in addressing these issues. In addition, it allows us to ascribe novelty scores to films on the basis of their content, including not just elements of the underlying story, but also elements that  encapsulate the tone and style of the final finished product. With this goal in mind, we analyze the plot-keywords associated with films produced in the United States over the period between and including the years 1890 and 2011, define two novelty scores based on them, and study the aggregate patterns in novelty evolution over a $70$ year period. In addition, we also provide a number of quantitative insights into the probability distribution of plot-keywords over the entire dataset spanning $100$ years, and the statistics of their use over time.

\section*{Results}
We begin by presenting some basic characteristics of the dataset under consideration. Henceforth for brevity, we will refer to plot-keywords simply as ``keywords''.

\subsection*{Statistics of films and tagging}

\begin{figure}[htbp!]
\includegraphics[width=180mm]{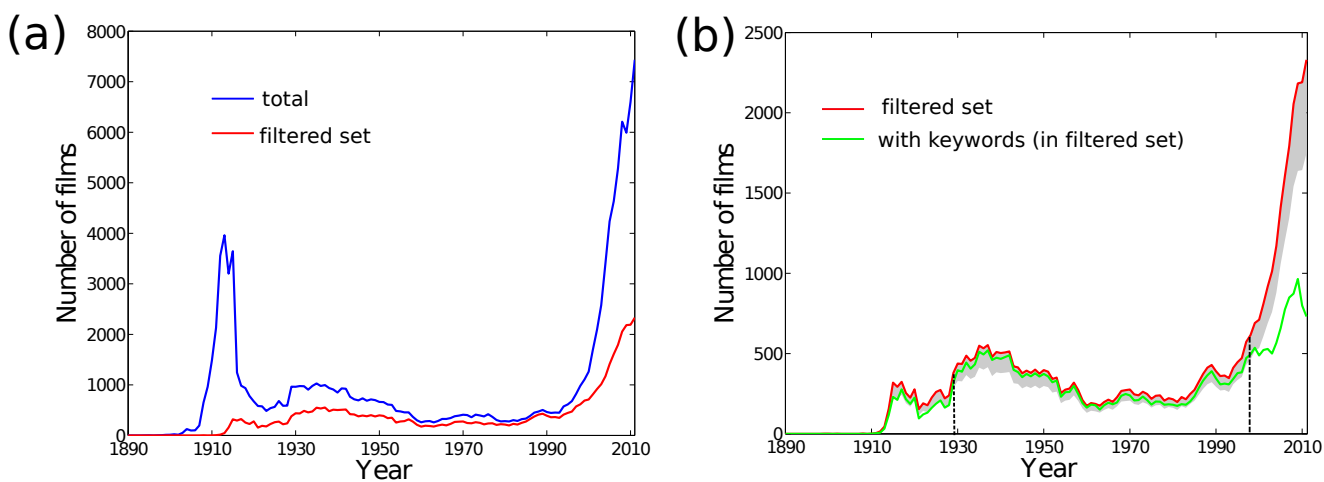}
\caption{(a) The total number of English language films produced in the United States (in blue), and the number of films remaining after filtering out short films, documentaries and adult films (in red), per year. (b) Number of films in the filtered set (red) and number of films in the filtered set with keywords (green), per year. The shaded gray region bounds the values which lie within $25\%$ of the total number of films released. In the period between and including the years 1929 and 1998 the green curve lies within the shaded region showing that greater than $75\%$ of films released each year in this period have keywords associated with them.}
\label{numberofmovies}
\end{figure}

Figure~\ref{numberofmovies}(a) shows the total number of English-language films originating in the US (see Methods for details) each year starting from the earliest recorded entry in the year 1890 through 2011. The number of films produced increases sharply starting around 1907, and corresponds to the ``Nickelodeon boom'' i.e., the sudden increase in the production of films as a result of the success of the Nickelodeon theater in 1905, which led to the proliferation of theaters devoted to film projection for a mass audience. The majority of the films produced in this period had runtimes of $10$-$15$ minutes \cite{historyoffilm}, and are classified as ``Short'' under the IMDb-Genre field. To obtain the dataset that forms the core of this study, we considered only feature length films, and additionally only those which were non-adult and non-documentary theatrical releases.  As expected the peak around 1910 disappears in the filtered set. Analogous to the Nickelodeon boom, there is a sharp rise in the number of films around the mid-1990s. This is a manifestation of the dramatic increase in independent-film production that occurred in the 1990s and that, by the end of the decade, led to over half the feature length films being produced coming from independent studios and producers \cite{contemporaryhollywood}. 

Figure~\ref{numberofmovies}(b) shows the statistics for the tagging of films released in the period between 1890 and 2011. Clearly, the association of keywords to films is not consistent over the different release years, with a clear paucity in tagging towards the early (the first film associated with a keyword was released in 1910) and late years in the period under consideration. However, for years in the period including and between the years 1929 and 1998, more than $75\%$ of the films released each year have keywords associated with them. For our studies on the novelty of films, we therefore focus on the films released within this period. In total, there are $21,583$ films possessing at least one keyword in this period. 

\begin{figure}[h!]
\includegraphics[width=180mm]{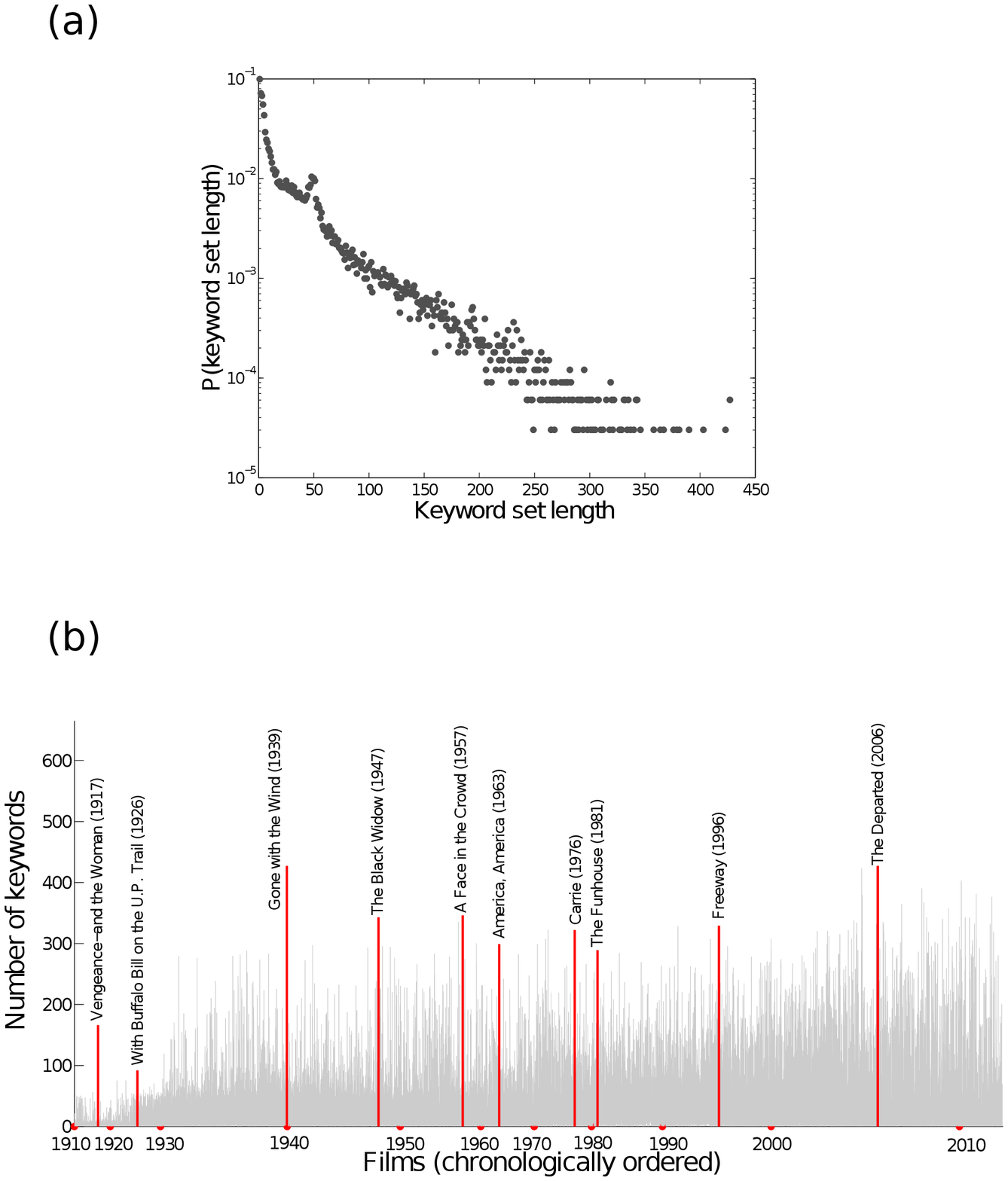}
\caption{(a) The distribution of keyword set lengths over all films with keywords. The linear decay on the linear-log plot indicates a roughly exponentially declining probability as the keyword set length increases.  (b) Length of the keyword set for the chronologically ordered set of films with keywords. The gray bars indicate the lengths of the sets for the different films. For each decade, the film with the longest keyword set over all releases in that decade is highlighted in red.}
\label{keywordlengths}
\end{figure}

We refer to the collection of all keywords associated with a film as the film's {\it keyword set}. The length of keyword sets appears to be exponentially distributed (see Fig.~\ref{keywordlengths} (a)), with the median length being $14$ keywords. For the restricted set between 1929 and 1998, the median length increases slightly to $19$, but the distribution remains qualitatively similar (not shown). As expected, films in the tail mostly comprise of popular mainstream films, as shown in Fig.~\ref{keywordlengths}(b) for each decade from the $1930$s to the $2000$s. 

\begin{figure}[h!]
\includegraphics[width=180mm]{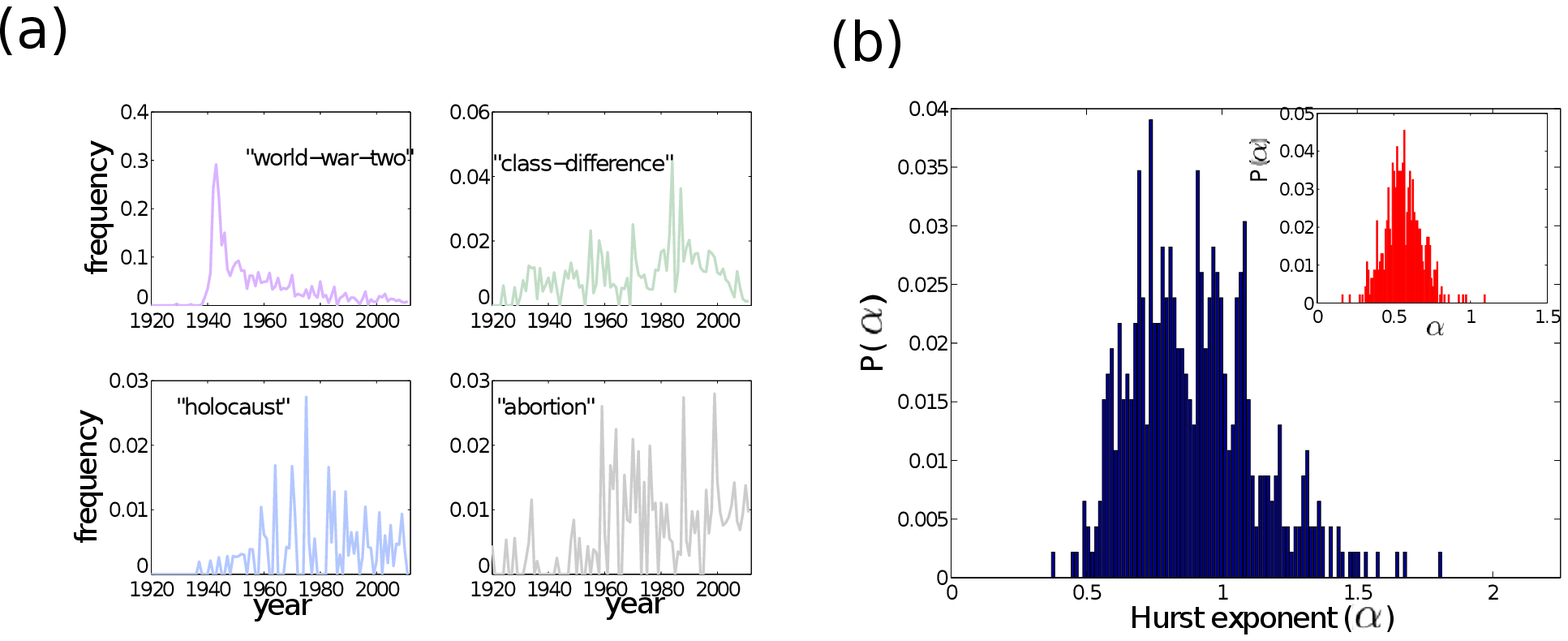}
\caption{(a) The yearly occurrence frequency of specific keywords as a function of time. (see text for details)(b) Distribution (relative frequencies) of the Hurst exponent $\alpha$ for keywords that occur in at least $70$ years between the period 1910 and 2011. The mean value of the exponent is $0.8966$, indicating the presence of positive long-range correlations. The inset shows the distribution after shuffling each of the time series. The correlations largely disappear upon shuffling as indicated by the mean value of $0.5590$ obtained for $\alpha$.}
\label{timeseries}
\end{figure}

Studies on the Google n-gram corpus have demonstrated that trajectories of word-occurrence-frequency over time can reflect surges of cultural interest in specific events, literary works, persons etc. \cite{Michel, Petersen}. 
We can expect to glean similar insights from observing the usage of plot-keywords.
We begin by defining occurrence frequency per year for a given keyword as the number of films released that year that are tagged with the keyword, divided by the total number of films released that year.  Figure ~\ref{timeseries}(a) shows trajectories of occurrence frequency for four example keywords. Similarly as observed for words in literature \cite{Michel,Petersen}, films too display a temporally local burst in the usage of a plot-element as can be seen in the example of ``world-war-two". A surge in the occurrence of ``class-difference" around 1985 is suggestively coincident with the conjectured rise in materialistic attitudes during the 1980s \cite{Collins,Materialism}. 

Beyond the temporally local trends seen in the association of keywords with films, there could also be long-range correlations present. To probe this further, we use the method of detrended fluctuation analysis (DFA) \cite{Peng} that is widely employed for investigating the presence of long-range correlations in general time-series, and has also been specifically used in the context of word usage \cite{Petersen}.  We analyzed using DFA (see Methods), the time series of keyword occurrence frequency for all keywords that appeared in at least $75$ of the years between the period 1910 - the earliest year with a tagged film - and 2011. In total, there are $461$ such keywords. The Hurst exponent $\alpha$ which signals the presence or absence of long range correlations is obtained for each of these time series using DFA. A value of $\alpha = 0.5$ indicates no temporal correlations, $\alpha < 0.5$ indicates negative correlations while $\alpha > 0.5$ indicates positive correlations. A distribution of the Hurst exponents obtained for the $461$ time series considered is shown in Figure~\ref{timeseries}(b), indicating the presence of long-range positive correlations in the keyword occurrence frequency. These correlations disappear (see Fig.~\ref{timeseries}(b) inset) for the set of time series obtained after shuffling the temporal order of data within each individual time series.

\subsection*{Evolution of film novelty}
Next, we devise a method to assign a novelty score to each film on the basis of the keywords associated with it and the keywords appearing in all films that were released prior to it.  The assignment of novelty scores is done for films in the continuous period between 1929 and 1998, more than $75\%$ of which per year are associated with keywords. In addition to the fact that keyword data is abundantly present for films released in or after 1929, the choice of this year as the beginning of our time window is also motivated by the fact that by then film-going was no longer an esoteric form of entertainment \cite{historyoffilm}, with film-ticket sales in the 1930s constituting as much as $4/5$ths of all entertainment expenditure \cite{Sedgwick}. Incidentally, the year 1929 marks the time around which sound in films became ubiquitous \cite{historyoffilm}, the beginning of the period which came to be known as the golden age of Hollywood \cite{goldenage}, and the year in which the first ever academy awards were presented. 
 We formally present the definition of the novelty score below.

For a film $i$, denote by $M^i$ the set of all films that appear prior to the release of film $i$. We use $m$ to index an arbitrary film, and $K_m$ to be the set of keywords associated with $m$. We begin by computing the probability $P(w)$ of observing a keyword $w$ over the set of films $M^i \cup \{i\}$ for all keywords appearing in the set.
\begin{equation}
 P(w) = \frac{1}{|M^i| + 1}\sum_{m \in M^i \cup \{i\}} \mathbbm{1}_{K_m}(w)
 \label{probkeyword}
 \end{equation}
 where $\mathbbm{1}_A$ denotes the indicator function for set $A$:
 \begin{equation}
 \mathbbm{1}_A(x) = \begin{cases} 1 &\mbox{if } x \in A \\
 							0 &\mbox{if } x \notin A \end{cases}
\end{equation}
Then, for any keyword $w$, the quantity -$\log P(w)$ is a standard measure of the ``surprise'' in observing keyword $w$ \cite{Cover}. With this in mind, we can quantify the novelty of film $i$, as the average surprise over all keywords associated with the film. Although, ideally, $P(w)$ should designate the prior probability distribution i.e., the probability distribution for keywords computed over films in $M^i$, we include film $i$ in its computation in order to circumvent the ill-defined logarithm arising when $P(w) = 0$ i.e., when $w$ appears for the first time in $K_i$. Thus, the first measure of novelty we define, aims to score the film on the basis of how rarely, on average, the elements associated with it have appeared in films in the past. For a given film $i$, the average surprise associated with its keyword set can be written as:
\begin{equation}
 \langle -\log P(w) \rangle = -\frac{1}{|K_i|}\sum_{w \in K_i}  \log P(w) 
\label{meansurprise}
\end{equation}
While formally appropriate as a measure of novelty, the above quantity suffers from the disadvantage that its maximum attainable value, $\log(|M^i|+1)$, is dependent on how many films have been released prior to the film under consideration. To yield a fair comparison between films irrespective of their position in the temporal order, we normalize the surprise associated with each keyword by the maximal attainable surprise ($\log(|M^i|+1)$ for film $i$), and define the {\it elemental novelty} associated with film $i$ as:
\begin{equation}
\mathcal{N}_{E}^{i} = -\frac{1}{|K_i|(\log(|M^i|+1))}\sum_{w \in K_i}  \log P(w) 
\label{elemental}
\end{equation}
Thus, $\mathcal{N}_E^{i}$ represents how close the average surprise for film $i$, as defined by Eq.~\ref{meansurprise}, is to its maximum attainable value.

While Eq.~\ref{elemental} scores films based on the rarity or abundance of their individual plot-elements, it is agnostic to how rare or abundant the {\it combinations} of their plot-elements are. To capture the novelty associated with the combinations of keywords, we can define similarly to Eq.~\ref{elemental}, the novelty resulting from the occurrence of specific keyword-pairs, triples and so on. Here we restrict our study of higher-order terms to keyword-pairs and formally write the {\it combinatorial} novelty for film $i$ as:
\begin{equation}
\mathcal{N}_{C}^{i} =  -\frac{1}{|K_i|(|K_i|-1)(\log(|M^i|+1))}\sum_{u,v \in K_i}  \log P(u,v) 
\label{combinatorial}
\end{equation}
where $P(u,v)$ is the probability of keywords $u$ and $v$ occurring together in a film in the set $M^i \cup \{i\}$ (defined similarly as for individual keywords in Eq.~\ref{probkeyword}).
Both $\mathcal{N}_E^i$ and $\mathcal{N}_C^i$ have values in the range $[0, 1]$, but capture distinct aspects of novelty generation.Thus observing trends in their evolution over time, not only gives us insights pertinent to specific events in the history of cinema, but also helps elucidate the degree to which elemental and combinatorial novelty contribute to the creation of new content.

\begin{figure}[h!]
\centerline{\includegraphics[width=150mm]{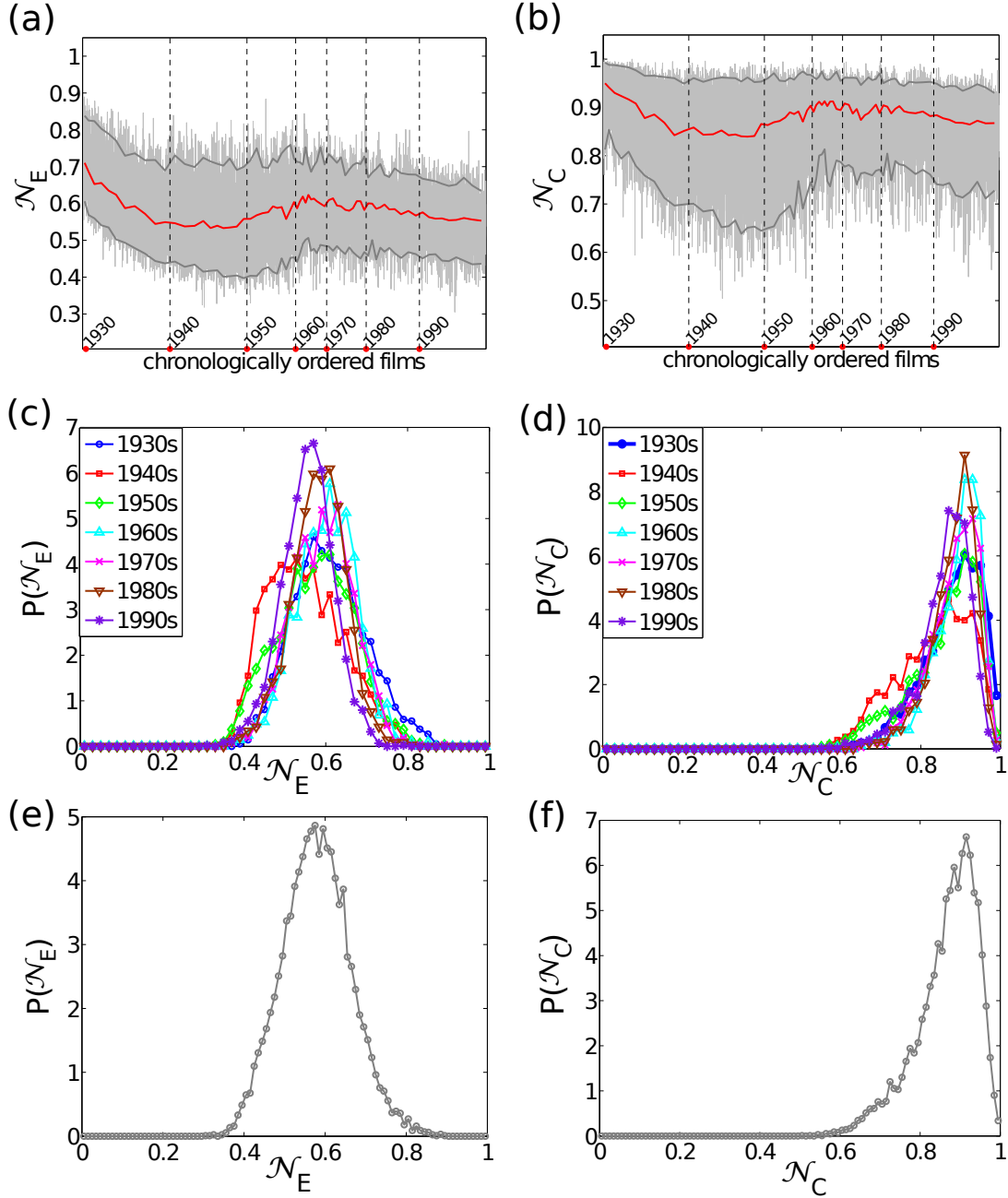}}
\caption{The evolution of (a) elemental novelty and (b) combinatorial novelty for films between 1920 and 1998.  The solid red curve shows the median yearly novelty, and the gray {\it envelope} curves show the novelty of the $5$th and $95$th percentile of films each year. Distributions of (c) elemental novelty and (d) combinatorial novelty by decade. Distribution of (e) elemental and (f) combinatorial novelty for the aggregated set of films released between 1929 and 1998.}
\label{Trends}
\end{figure}

%
Figure~\ref{Trends}(a) shows the chronological evolution of elemental novelty over the period 1929-1998. To eliminate situations where a film with a small keyword set registers a very high (very low) novelty due to the rarity (abundance) of its few keywords, we only consider films with keyword sets of length greater than $10$ (see SI Section 1.1).  Films are chronologically ordered by the time of release, and the abscissa is simply the index $i$ of the films, with the vertical dashed lines corresponding to the indices demarcating the beginning of a new decade.As stated earlier, novelty is bounded above by $1$, and the median value of elemental novelty (shown in red) is well below this bound over the entire period. Some features in the evolution also bear pointing out. For example, an upward trend can be seen around the mid-1960s in both the yearly median, as well as the lower envelope of the time series, which agrees well with the documented birth of the American New Wave which brought with it a marked shift in themes, style and modes of production\cite{historyoffilm}. Interestingly, the period between 1929 and 1945, commonly referred to as the golden age of Hollywood, is not marked by an increase in or a stable value of median novelty, but rather by a subtle decline. This decline is likely a consequence of the practice of block booking prevalent in that period, which by virtually guaranteeing exhibition for any film as long as it came from a major studio, did little to de-incentivize the production of films with low novelty \cite{Devanybook,historyoffilm}. 

Figure~\ref{Trends}(b) analogously shows the evolution of combinatorial novelty over the period, whose upper envelope in contrast to elemental novelty, consistently stays close to the maximum attainable value of $1$. Gross features similar to those seen for elemental novelty can also be seen here; the median $\mathcal{N}_C$ rises in the 1960s and its variance decreases, while in contrast, the variance shows an increasing trend during the ``golden age'' between 1929 and 1945.

Figure~\ref{Trends}(c) and (d) respectively show the probability density functions (pdf) of elemental-novelty and combinatorial-novelty for films in each of the $7$ decades in the period considered. All the distributions are unimodal, differing slightly in their variances, but with their respective modes confined to the range between $0.53$ and $0.63$ for $\mathcal{N}_E$, and between $0.87$ and $0.93$ for $\mathcal{N}_C$. For each type of novelty, the similarity between individual decade-wise pdfs and the overall pdfs, Figs.~\ref{Trends}(e),(f) respectively, hint at the possibility of some underlying novelty preferences governing which scripts are chosen for development into a feature film. 


\begin{figure}[h!]
\centerline{\includegraphics[width=150mm]{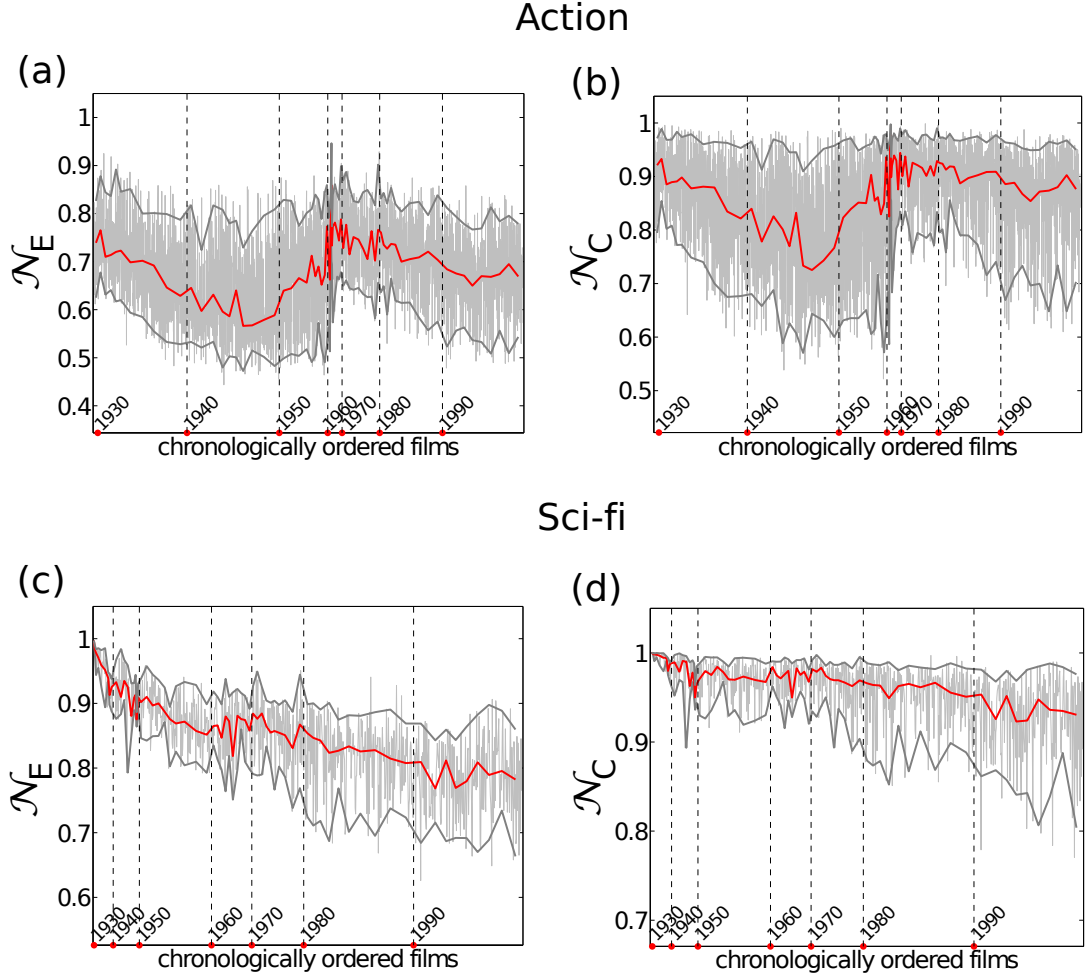}}
\caption{(a) Elemental novelty and (b) combinatorial novelty for films containing `Action' within their `genre' field on IMDb.  (c) Elemental novelty and (d) combinatorial novelty for films containing `Sci-Fi' within their `genre' field on IMDb. The solid red curve shows the median yearly novelty, and the gray envelope curves show the novelty of the $5$th and $95$th percentile of films each year.}
\label{Trends_genre}
\end{figure}

We also investigate the evolution of elemental and combinatorial novelty for films within specific genres, and these reveal trends unique to each of them.
For example, Fig.~\ref{Trends_genre}(a), (b) show the evolution of novelties for films containing ``Action'' as one of their IMDb genre classes while Fig.~\ref{Trends_genre}(c) and (d) show the case for films under the ``Sci-fi" genre. 
The median and the envelope curves of both $\mathcal{N}_E$ and $\mathcal{N}_C$ for the case of action films, show a sudden disruptive jump to higher values in the decade 1960-70. This is compatible with the thesis, based on studies by film historians, that elements comprising the modern action film genre originated with the James Bond franchise in the 1960s \cite{Bond}. Similar plots for selected other genres are shown in Supplementary Figure 2.

\subsection*{Relationship between film novelty and revenue}


Next, motivated by the Wundt-Berlyne curve, we investigate whether there is any relationship between the novelty of a film and the hedonic value derived from its consumption at an aggregate population level. 
We utilize the (inflation-adjusted) revenue generated by the film as a measure of its mass appeal (see Supplementary Text, Section 1.2), and measure a film's novelty only taking into account the films released in a $6$ month window prior to its release (see Methods for details).

\begin{figure}[h!]
\centerline{\includegraphics[width=140mm]{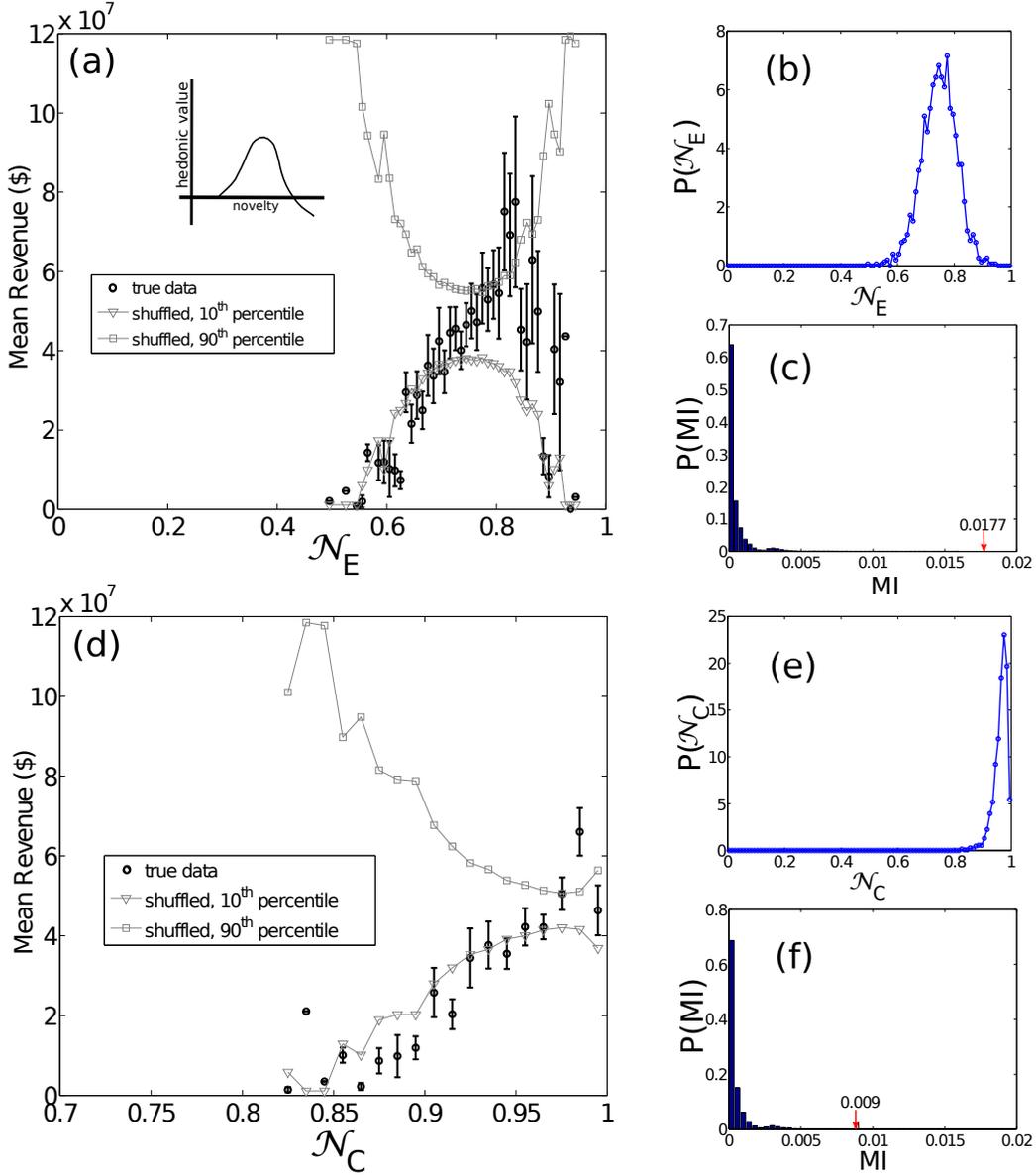}}
\caption{(a) Mean inflation-adjusted revenue versus elemental novelty $\mathcal{N}_E$ is shown by the black circles, with vertical segments indicating standard-errors in the computed mean values. Also shown are the $10^{th}$ (gray triangles) and $90^{th}$ percentile (gray squares) of mean revenues obtained for $50000$ randomized versions of the data. (b) The probability density function of $\mathcal{N}_E$ for the data used in (a). (c) The relative frequencies of mutual information values between $\mathcal{N}_E$ and mean revenue obtained for the randomized datasets, compared to the mutual information for the true dataset. (d) Mean inflation-adjusted revenue versus combinatorial novelty $\mathcal{N}_C$ (black curve) and the $10^{th}$ and $90^{th}$ percentile (gray triangles, gray squares respectively) of mean revenues obtained for $50000$ randomized versions of the data. (e) The probability density function of $\mathcal{N}_C$ for the data used in (d). (f) The relative frequencies of mutual information values between $\mathcal{N}_C$ and mean revenue obtained for the randomized datasets, compared to the mutual information for the true dataset.}
\label{wundt}
\end{figure}

In Figure~\ref{wundt}(a) we plot the mean revenue of films conditioned on elemental novelty (black circles). The overall shape of the resulting curve shows a resemblance to the Wundt curve (inset), with the mean revenue increasing systematically with novelty until a value of around $0.8$, and declining thereafter. To get a better sense of the significance of this curve, we generate $50000$ randomized versions of data where the values of revenues are shuffled. For every shuffled data set we obtain the mean revenue corresponding to each novelty bin, and then plot the $10^{th}$ and $90^{th}$ percentile of all mean revenues values obtained for each novelty bin (gray curves). A significant fraction of the true data points either straddle these curves, lie below the $10^{th}$ percentile curve, or lie above the $90^{th}$ percentile curve, indicating that their respective probabilities of occurring purely due to chance is $\leq 10 \%$. The declining portion of the curve is harder to conclusively argue for, due to a paucity of data points for the associated range of novelty, as evidenced by the probability density of novelty, Fig.~\ref{wundt}(b). Irrespective of the precise nature of the relationship between novelty and hedonic value, we can investigate whether these two quantities exhibit a significant statistical dependance on one another. We do this by evaluating the Mutual Information (MI) (see Methods) between the two quantities, and comparing it to the values obtained from a permutation test. Specifically, we generate $50000$ datasets where the revenue values are shuffled, and compute the MI between novelty and revenue for each shuffled dataset. Figure~\ref{wundt}(c) shows that the MI for the true data (red arrow) is far from the tail of the distribution of MI values obtained using the shuffled datasets. More precisely, none of the shuffled datasets achieved a value equal to or greater than the true MI of $0.0177$, indicating a p-value less  than  $2\times10^{-5}$. 

Figure~\ref{wundt}(d) shows the mean revenue of films as a function of their combinatorial novelty. Here the range of novelty values is much narrower (Fig.~\ref{wundt}(e)), and the only discernible feature is a systematic increase in mean revenue as novelty increases. The MI between novelty and $\mathcal{N}_C$, as in the case of $\mathcal{N}_{E}$, is statistically significant as indicated by a permutation test, with a p-value less than $2\times10^{-5}$ (Fig.~\ref{wundt}(f)).

\subsection*{Overall occurrence probabilities of keywords and keyword-pairs}

  \begin{figure}[h!]
\includegraphics[width=150mm]{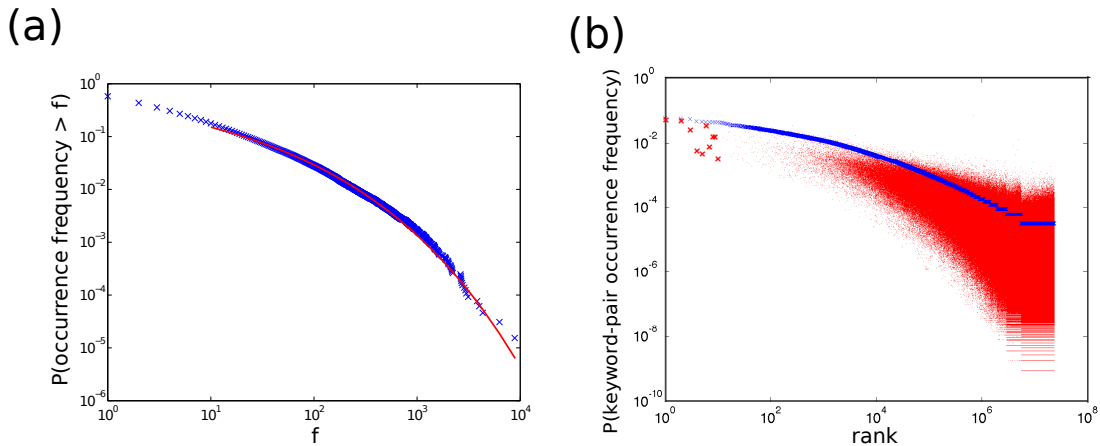}
\caption{(a) The cumulative probability, $P({\rm occurrence ~frequency} > f)$ for keywords. The red line shows a fit corresponding to the cumulative probability for a stretched exponential distribution, $\exp(-\lambda \;f^\beta )$ with parameters $\lambda = 1.0119$ and $\beta = 0.2716$. (b) The frequency of keyword-pair occurrence as a function of the rank of the pair (blue). For the keyword-pair corresponding to each rank, the probability of occurrence under the independence assumption is shown by a red dot. For the $10$ highest ranked keyword-pairs, the probabilities of occurrence under the independence assumption are indicated by red crosses. }
\label{probdist}
\end{figure}

Next, we study the probability distribution of plot-keywords over the entire set of films in the period between 1890 and 2011. Unlike the case for other corpora \cite{Michel,Petersen}, the distribution does not follow Zipf's law as seen from the curvature present in the log-log plot of the cumulative probability distribution of usage frequency (Fig.~\ref{probdist}(a)). Indeed, a stretched exponential fit obtained through maximum-likelihood-fitting \cite{Virkar, Clauset} agrees well with the data (parameters provided in caption). 


Any non-trivial process of plot generation would result in some keyword-pairs occurring more often than expected by chance, and others less often. To probe whether this is indeed borne out by the data, we compare the occurrence frequency of keyword pairs to the frequency obtained under the assumption that the constituent keywords are chosen independently of each other, in proportion to their respective occurrence probabilities. The results shown in Fig.~\ref{probdist}(b) show a substantial difference between the true keyword-pair frequencies and those obtained under the independence assumption.

\begin{figure}[h!]
\centerline{\includegraphics[height = 12cm]{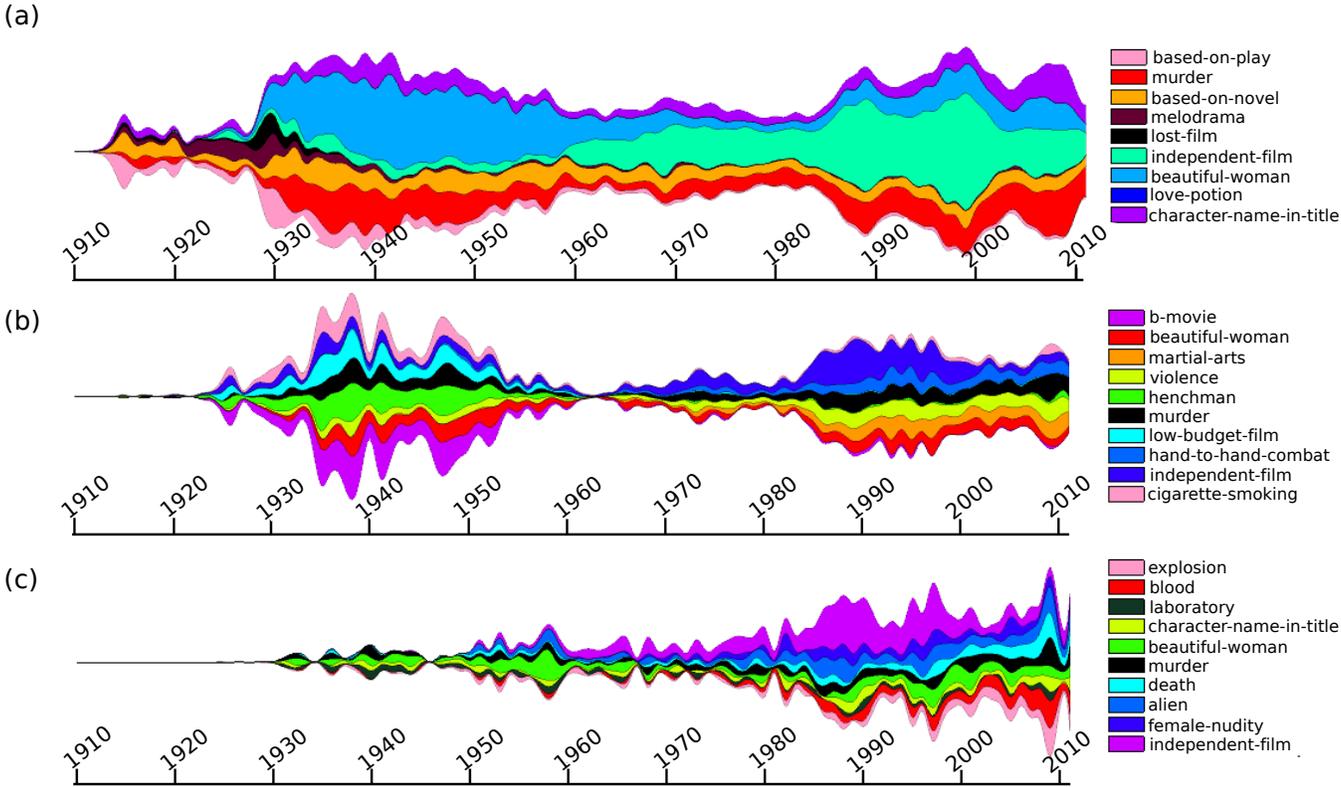}}
\caption{Streamgraphs for most probable keywords occurring in (a) all films (b) action films and (c) science-fiction films. See Methods for details.}
\label{streamgraphs}
\end{figure}

Finally, we present a visual depiction(Fig.~\ref{streamgraphs}) of the rise and fall of keywords that are associated with movies over the entire period from 1910 to 2011. Unlike a traditional time series plot (as in Fig.~\ref{timeseries}(a)) {\it streamgraphs} introduced in \cite{Havre, Byron} provide a lucid graphical approach to simultaneously observing the growth and decline in the usage of different keywords (thickness of each ``stream''), along with their relative usage in a given year (relative thickness of a stream in a cross section).

A prominently visible feature in Fig.~\ref{streamgraphs}(a) is the growth in the use of the keyword {\it independent-film} beyond 1955, presumably resulting from the demise of the studio system and marking the period when studios began forming partnerships with independent producers. Furthermore, until that time, the monopoly of the studios on the exhibition venues, strongly suppressed the visibility of independently produced films \cite{historyoffilm}. 
%
A notable feature in the action streamgraph (Fig.~\ref{streamgraphs}(c)) is the early dominance of the keyword {\it b-movie} and its decline in the 1950s. Indeed, between 1930 and 1950, action films mostly comprised of low-budget westerns created to fit the double feature programming format \cite{Nachbar}. However, by the 1950s, with film audience numbers in decline as a result of the predominance of television, and with the end of the studio-system, the low-budget action film gradually declined in production and the genre as a whole underwent a redefinition in the 1960s \cite{Bond}.

\section*{Discussion}
We have demonstrated that user-generated keywords coarsely characterizing a film, can provide a quantitative window into the evolution of novelty in films over a $70$ year period. Specifically, the novelty scores defined here reveal both subtle trends in overall novelty evolution (Fig.~\ref{Trends}) and disruptive changes in the evolution of specific genres (Fig.~\ref{Trends_genre}(a)). A notable feature of several evolution curves is an upward trend in novelty during the 1960s (Fig.~\ref{Trends}(a),(b),  and Fig.~\ref{Trends_genre}(a),(b)). Presumably, this corresponds to the widely held thesis \cite{historyoffilm} that the break-up of the studio system, the advent of competition from television, and the rise of several socio-political movements, all contributed in varying measures to the 1960s becoming a defining decade in the history of American cinema. 

However, the fact that the overall distributions as well as the decade-wise distributions of $\mathcal{N}_E$ and $\mathcal{N}_C$ overlap significantly, suggests some strong constraints on the degree of novelty in films that eventually get made and released theatrically. This could be a manifestation of the inherent novelty preferences of the investors, or of risk-minimization based on some implicitly perceived inverted-U relationship between novelty and hedonic value. Indeed, the plot of $\mathcal{N}_E$ versus mean-revenue, Fig.~\ref{wundt}(a) does lend some credence to the idea that the relationship between novelty and hedonic value resembles the Wundt-Berlyne behavior.


While this study has focussed on utilizing keywords to observe aggregate trends, there are several possible extensions that can be pursued in future work. The first is to attempt a refinement of novelty scores which takes into account the descriptive level of the keyword, an issue that is ignored in this study. For example, here we treat a keyword characterizing a high-level feature related to the production (for example {\it independent-film}) equivalently to a keyword which specifies a story-element (for example {\it murder}). A possible approach to alleviating this is by employing a probabilistic topic model like hierarchical latent Dirichlet allocation on the keyword set \cite{Blei}, and then defining a more finely resolved measure of novelty based on the obtained hierarchy of topics.
 
A second potential research direction is to analyze the utility of the novelty score discussed here or refinements of it to search and recommendation. Yet another application of such scores is in the area of artificial or computer-aided story generation \cite{Perez} where ranking the novelty of plot-element combinations based on their prior probabilities could allow exploration in novel directions. Understanding aggregate novelty preferences may also provide insights into the viral spread and mass adoption (or lack thereof) of certain products and services, and is a research direction with valuable applications to marketing campaigns and social network based behavior-change initiatives. Furthermore, any venue offering the combined availability of crowdsourced data, the network between users providing tags, and their individual tagging behavior, provides the opportunity to segment the population on the basis of their novelty preferences, and design products and services tailored specifically to each segment.

\section*{Methods}
\subsection*{Data collection and analysis:}
Data was obtained from IMDb (http://www.imdb.com/interfaces) as plain text data files in May 2012. Data was processed with Python scripts using the IMDbPY package (http://imdbpy.sourceforge.net/). First, all data items corresponding to films (not including straight-to-video releases, or TV movies) were extracted. Next, those items which had ``Country" listed as `USA' and ``Language" listed as `English' were extracted. Finally, all films with `Adult', `Short' or `Documentary' under ``Genre" were removed to leave us with the set under consideration. For more details, see Supplementary Text, Section 1.1.

\subsection*{Detrended fluctuation analysis:}
Detrended fluctuation analysis for a time series $y \equiv \{y_1,y_2,\cdots y_N\}$ involves the following steps:
\begin{itemize}[]
\item{(i)} Mean-center the original time series:  $\bar{y} \equiv \{y_1-\langle y \rangle, y_2 - \langle y \rangle, \cdots, y_N -\langle y \rangle \}$ where $\langle y \rangle = \frac{\sum_{i=1}^{N} y_i}{N}$
\item{(ii)} Generate a random walk $z$ by summing up displacements corresponding to values in $\bar{y}$:  $z_j = \sum_{i=1}^{j} \bar{y}_i$
\item{(iii)} Partition the total number of steps in the walk (i.e., total number of elements in the original time series) into boxes of size $L$. 
\item{(iv)} Within each box, compute the local trend $\bar{z}$ using a linear fit to the data. Compute the variance in the detrended fluctuations within each box and then compute the square root of its average over all boxes: $\sigma(L) \sim \sqrt{\langle \overline{ (z(t)-\bar{z})^2} \rangle}$ (where the $\langle \cdots \rangle$ corresponds to an average over boxes, and the term within corresponds to the variance within a box. 
\item{(v)} Repeat the process for different values of $L$ and estimate the exponent $\alpha$ in the scaling $\sigma(L) \sim L^{\alpha}$.
\end{itemize}

\subsection*{Mutual Information estimation:}
The mutual information between random variables $x$ and $y$ with marginal distributions $P(x)$ and $P(y)$ respectively and joint distribution $P(x,y)$ is defined as:
\[
I = \sum_{x,y} P(x,y) \log\bigg(\frac{P(x,y)}{P(x)P(y)}\bigg)
\] 
In the absence of a knowledge of a specific form for the relationship between variables, mutual information is a useful signifier of the presence or absence of dependencies between variables $x$ and $y$ \cite{Steuer}. The estimation of mutual information between two continuous variables with a finite number of observations is a well-studied problem. We utilize a method proposed in \cite{Vajda, Darbellay} and an implementation of the same provided by Zbynek Koldovsky.

\subsection*{Novelty and hedonic value}
The following pertains to the data and methods used for Fig.~\ref{wundt}. Budgets and revenues generated from theatrical exhibition are present for $1680$ films in the period under consideration. We adjust for inflation all dollar amounts that have a reporting year associated with them based on the cumulative price index table for the year 2011. To strike a balance between having a sufficiently large number of films to analyze, and minimizing the disparities in the exhibition capabilities of films considered, we restrict our analysis to films with a inflation adjusted budget of at least $1$ million dollars (see SI, Section 1.2 for further details). Finally, to account for the fact that novelty as perceived by a general audience largely involves comparison to films released over a short period in the past (rather than the over the entire duration that cinema has been around), we compute $\mathcal{N}_E$ and $\mathcal{N}_C$ for a film $i$, only considering films which were released in the $6$ months preceding the month of its release.

\subsection*{Streamgraphs}
A ``stream" for a keyword was generated using the number of occurrences of the keyword for each year in the period. The resulting signal was smoothed using spline interpolation. A stacked graph was generated and to guarantee symmetry about the Y axis, the baseline was displaced in proportion to the total width of the stack as described in \cite{Havre,Byron}.
For Fig.~\ref{streamgraphs}(a) we use the set of keywords obtained from the union of the most frequently used keyword for each year in the period. This set contains $9$ unique keywords. For streamgraphs shown in Figs.~\ref{streamgraphs}(b) and (c) for films belonging to the action and science-fiction genres respectively, keyword sets were chosen using a similar procedure as for Fig.~\ref{streamgraphs}(a) but were additionally pruned to retain only the $10$ keywords with the highest average usage-frequency over the period. 

\section*{Acknowledgments}
This work was supported in part by the Army Research Laboratory under Cooperative Agreement Number W911NF-09-2-0053 and by the Office of Naval Research Grant No. N00014-09-1-0607. The views and conclusions contained in this document are those of the authors and should not be interpreted as representing the official policies either expressed or implied of the Army Research Laboratory or the U.S. Government. 
S.S. thanks B. K. Szymanski, G. Korniss and A. Asztalos for critical readings of the manuscript, and valuable comments and suggestions. S.S. thanks Y. Virkar for a Matlab implementation of the maximum-likelihood-fit of a stretched exponential function to binned data, and Zbynek Koldovsky and  Petr Tichavsky for a Matlab implementation of the estimation of mutual information using adaptive partitioning.

\section*{Author Contributions}
S.S. designed the research, performed the analysis of data and wrote the manuscript.

\section*{Additional Information}
The authors declare no competing financial interests.



\renewcommand{\figurename}{Supplementary Figure}
\setcounter{figure}{0}

\newpage

\section{Supplementary Text}

\subsection{Details of the dataset used}

Data from the Internet Movie Database was obtained as plain text files from the Alternative Interfaces page: (http://www.imdb.com/interfaces).
Data was downloaded in May 2012. Only data pertaining to films through 2011 was used.
Data was processed using the Python IMDbPy module: (http://imdbpy.sourceforge.net/).
We first extracted all films that contained `USA' in the field `Country'. This corresponded to all films produced (or at least attributed to) the United States.
Next, from the above set we extracted films that did not contain the terms `Short', `Adult' or `Documentary' in its `Genre' field.
The number of films in this filtered set is $46596$. The total number of films with keywords in this set is $33128$ and the earliest film with a keyword has a release year of 1910.

For Figs. 1, 2, 3, 7 and 8 in the main text, wherever keywords are considered, data from all films between 1910-2011 was used.

For Figs. 4, 5 and 6 in the main text, novelty was calculated for films released between 1929 and 1998, the period for which there is a fairly consistent degree of tagging (see Fig. 1(b) of main text). For Figs. 4 and 5, when computing the novelty of a film $i$, the probability of keyword usage $P(w)$ was calculated using all films with keywords that were released prior to it (the earliest being in 1910), as well as the film under consideration. For Fig. 6, when computing the novelty of a film, the probability of keyword usage was calculated using all films that appeared $6$ months prior to the release month of the film, and the film under consideration. 

Furthermore, for Fig. 6, only films with an inflation adjusted budget greater than or equal to \$ 1 million were considered (see next section for explanation).  

In Figs. 4,5 and 6, the films were additionally filtered to retain only those with greater than $10$ keywords in their respective keyword sets. Eventually, the total number of films used for the results in Fig. 4 was $13322$, which constitutes approximately $62\%$ of all films with keywords in the period 1929 to 1998. The total number of films used for Fig. 6 after filtering for keyword set lengths and budget was $1509$. 

\subsection{Choosing a proxy for hedonic value}

\begin{figure}[h]
\begin{center}
\includegraphics[width = 6in]{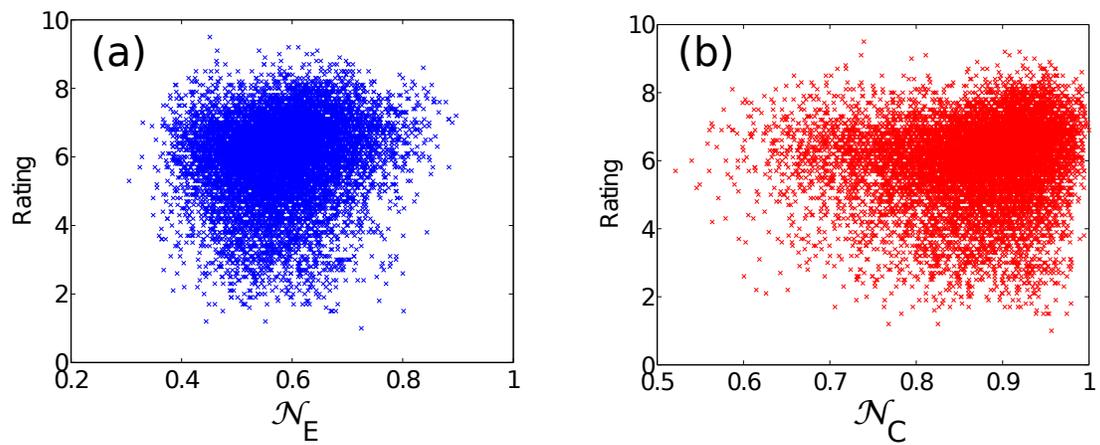}
\end{center}
\caption{Scatterplot of (a) $\mathcal{N}_E$ and (b) $\mathcal{N}_C$ versus IMDb rating for all films (with keywords) between 1929 and 1998.}
\label{rating}
\end{figure}

The IMDb dataset provides two possible avenues to capture a proxy for the hedonic value derived from a film. The first of these, the IMDb rating, is a weighted average of individual user ratings, obtained using an undisclosed weighting scheme that is designed to counter ballot-stuffing \cite{IMDBlink}. How effectively the ratings alleviate the problem is unclear, but even in the best-case scenario when voting is honest, the rating would reflect the taste preferences of only the registered users of IMDb. Perhaps, as a consequence of the idiosyncrasies in the computation of IMDb ratings, the relationship between novelty and rating shows scarce decipherable structure (see Fig.~\ref{rating}). More pertinently, even for a film that was released after IMDb was established, we only have access to its current (i.e. at the time when the data was downloaded) IMDb rating, and not its rating immediately after its release, which would accurately reflect how it was received. 
Since our measures, Eqs.~4 and 5 in the main text represent the novelties of a film specifically at the time of its release, for a principled investigation of the relationship between novelty and hedonic value, we require a quantity that captures the perceived value of the film precisely around the time of its release. This requirement clearly renders the IMDb ratings that we have access to, unsuitable for our purposes as a proxy for hedonic value. 
The second quantity available to us, and which we utilize here to represent the hedonic value, is the revenue generated by the film through theatrical ticket-sales. Specifically, we extract the revenue generated from ticket sales within the United States for films (whenever available), and adjust these values for inflation using the consumer price index table for 2011. This measure is much more appropriate for our purposes, since the theatrical run of the film is typically confined to within four months after the film's release \cite{nato}, and therefore yields a suitable proxy for hedonic value obtained over a time window proximal to the time of its release. 

%

One of the drawbacks of utilizing revenue (also rating) is its inability to represent
distaste or negative hedonic values. It might appear that using a measure like return on investment (ROI) 
which is the profit (loss) divided by the production cost might alleviate this shortcoming. 
However, we argue that any individual or aggregate measure of hedonic value should be agnostic to the production cost of the film. 
In other words, given two films with equal viewership (i.e., equal ticket sale revenue), it is unreasonable to ascribe a higher reward to the film with the lower budget, simply on account of its lower production cost.

Another possible argument in favor of incorporating budgets in the measure of reward is to counter the influence of budgets in drawing audiences; expensive films invariably have more theaters exhibiting them, thus suggesting that the production budget has a direct influence on viewership. However, an expensive film that performs poorly is liable to be removed from exhibition after the initial commitment period (the minimum contractually obligated period for which a theater screens the film) has elapsed, and a relatively inexpensive film playing in only a few theaters could outperform it in terms of viewership if it garners sustained audience interest. Thus the relationship between production cost and viewership is not straightforward. However, in order to mitigate issues which arise due to such differences in exhibition capability we only use films that have (inflation adjusted) budgets above \$1 million to obtain the results shown in Fig. 6 of the main text.
\newpage 
\section{Supplementary Figures}

\begin{figure}[h!]
\begin{center}
\includegraphics[width = 6in]{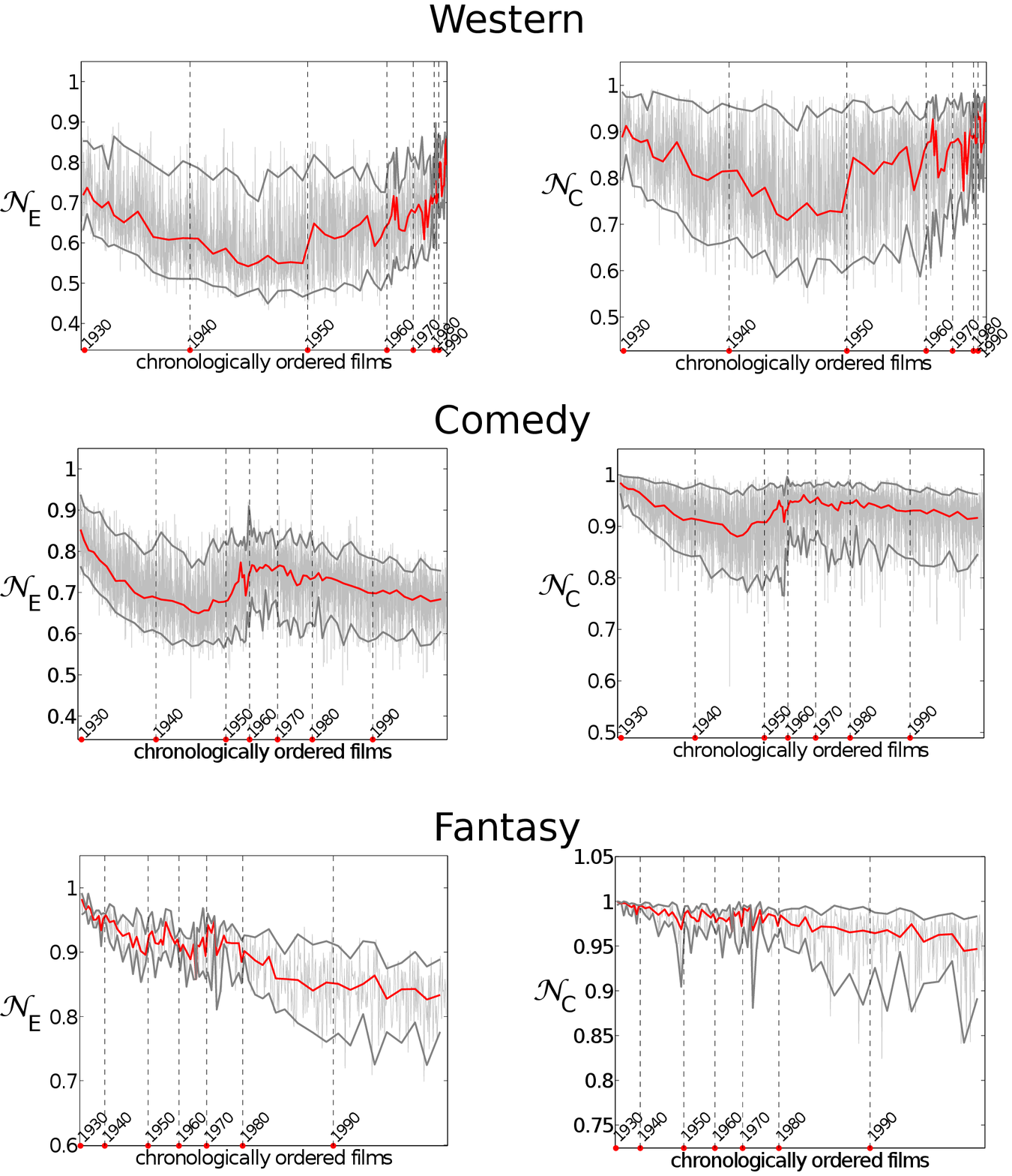}
\end{center}
\caption{Trends for films based on appearance of different terms in their IMDb `Genre' field.}
\label{othertrends}
\end{figure}

\begin{figure}[h!]
\begin{center}
\includegraphics[width = 6in]{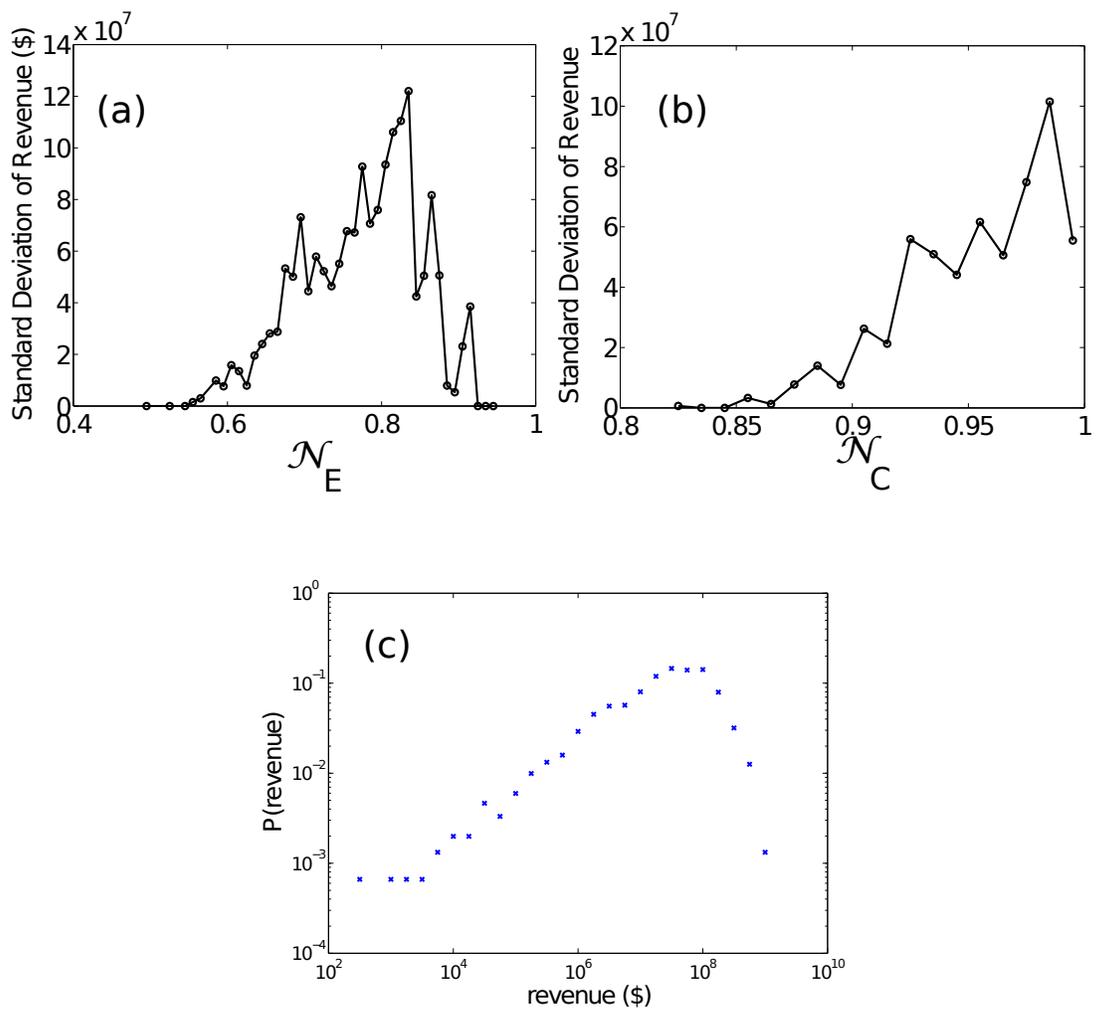}
\end{center}
\caption{Standard deviation of revenue as a function of (a) elemental novelty and (b) combinatorial novelty. (c) Probability density function of revenue for films considered in results of Fig. 6.}
\label{stdev}
\end{figure}

\begin{figure}[h!]
\begin{center}
\includegraphics[width = 4.5in]{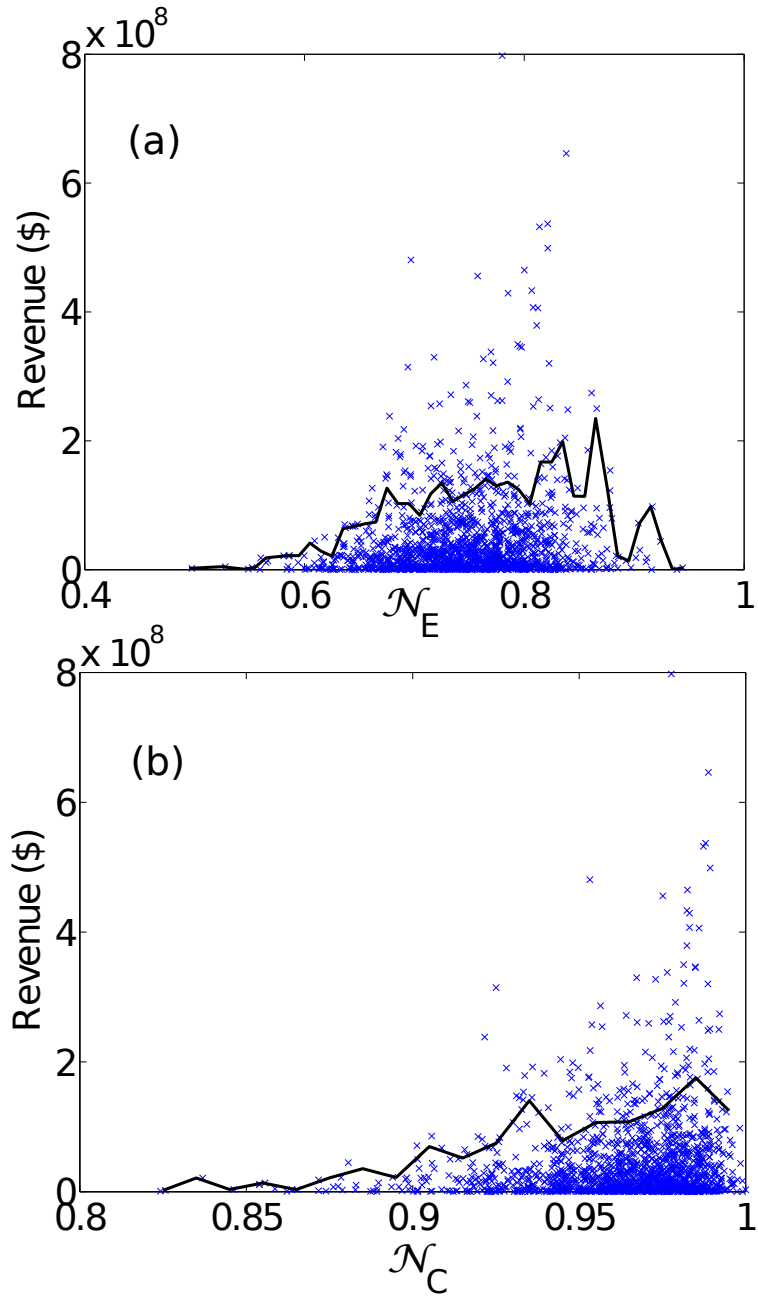}
\end{center}
\caption{Scatterplot of revenue as a function of (a) elemental novelty and (b) combinatorial novelty from the data used for Fig. 6 in main text. The black curve on both plots indicate the $90^{th}$ percentiles of revenue for each binned value of novelty. Novelty values were binned over the interval shown into $100$ bins.}
\label{scatter}
\end{figure}

\end{document}